\documentstyle[12pt,epsfig]{article}                                
\begin{document}

\titlepage

\begin{large}
\begin{bf}
Asymmetric factorization method on supersymmetry : Complex operators
\end{bf}
\end{large}

 Biswanath Rath

\vspace{0.1cm}

Department of Physics,
 Maharaja Sriram Chandra Bhanj Deo University,
 Takatpur, Baripada -757003, Odisha, INDIA

(*biswanathrath10@gmail.com)

\vspace{0.1cm}

$\bf{Abstract:}$
We propose asymmetric factorization method for supersymmetry involving complex operators. Model Hamiltonians
satisfy the supersymmetric energy conditions $E_{n+1}^{(-)}=E_{n}^{(+)}$; $E_{0}^{(-)}=0$.

\vspace{0.1cm}

\begin{bf}
PACS no-03.65.Ge,11.30.Pb
\end{bf}

\vspace{0.1cm}

\begin{bf}
Key Words :
\end{bf}

susy energy conditions,complex operators,PT-symmetry,non-self adjoint operators

\begin{bf}
1. Introduction
\end{bf}
Supersymmetry is a powerful technique for realizing real spectra,which has been 
highly studied under real operators[1,2], where susy Hamiltonians ($H^{\pm}$) are generated via superpotential ($W$) satisfying the relations
\begin{equation}
H^{-}=p^{2}+W^{2}-\frac{dW}{dx}
\end{equation}
and 
\begin{equation}
H^{+}=p^{2}+W^{2}+\frac{dW}{dx}
\end{equation}
The corresponding energy levels satisfy the relations
\begin{equation}
E_{n}^{(+)}=E_{n+1}^{(-)}
\end{equation}
and 
\begin{equation}
E_{0}^{(-)}= 0
\end{equation}
Further, we have 

\begin{equation}
H =
\left[
\begin{array}{cc}
H^{(-)} & 0  \\
0  & H^{(+)} \\
\end{array}
\right ]
\end{equation}
where any suitable form of $H^{(\pm)}$ may be considered.However, the same can has not been formulated using complex operators[1,2].Further, complex operators
must satisfy $PT$-symmetry condition[3].
Here, $P$ stands for parity operator having the behavior:
$PxP^{-1}=-x; P p P^{-1}=-p$. 
Similarly,T stands for the time reversal operator having the 
following operations: $TxT^{-1}=x;  T p T^{-1}=-p $. Now we formulate susy Hamiltonians for complex operators as follows. 

\begin{bf}
2. Explicit form of $H^{(-)}$ and $H^{(+)}$
\end{bf}
  
We ensure that operator $A$ be annihilation operator  satisfying the condition
[4]
\begin{equation}
A= i\frac{d}{dx} + W_{1}(x)
\end{equation}
where 
\begin{equation}
[PT, W_{1}]=0
\end{equation}
Similarly the similar behaviour of $W_{2}$. Hence it is clear that
\begin{equation}
[W_{2}, W_{1}]=0  
\end{equation}
Hence using $W_{2}$, we define another operator $B$ such that 
\begin{equation}
B= i\frac{d}{dx} + W_{2}(x) 
\end{equation}
with $W_{1}(x)\neq W_{2}(x)$. 

Now using $A,B$ we define $H^{(\pm)}$ as [4]
\begin{equation}
H^{(-)}= BA = p^{2} - W_{1}\frac{d }{i d x} -\frac{d W_{1}}{id x}- W_{2}\frac{d}{idx} + W_{1}W_{2}
\end{equation}
and 
\begin{equation}
H^{(+)} = BA = p^{2}- W_{2}\frac{d }{id x}-\frac{d W_{2}}{i d x} - W_{1}\frac{d}{idx} + W_{1}W_{2}
\end{equation}
Under this case the ground state wave function is written as 

\begin{equation}
\psi_{0}\sim e^{-\int{\frac{ W_{1}}{i}} dx}
\end{equation}
 Similarly, we have 
\begin{equation}
\psi^{+}_{n}(x) = (E_{n+1}^{-})^{-1/2} A \psi_{n+1}^{-}(x)
\end{equation}
\begin{equation}
\psi^{-}_{n+1}(x) =  (E_{n}^{+})^{-1/2} B \psi_{n}^{+}(x)
\end{equation}
It should be borne in mind that $W_{1,2}$ will have explicit time reversal nature. 
In order to show this below we consider a few explicit functions of $W_{1} ; W_{2}$. So that the product$W_{1} W_{2}$ will generate complex terms.

\begin{bf}
3.Model  example on $W_{1}$ and $W_{2}$. 
\end{bf}

Here we consider different choices of $W_{1}$ and $W_{2}$ as follows.

\begin{bf}
Example-1:Unequal terms 
\end{bf}

Let us consider unequal terms in $W_{1}$ and $W_{2}$.Here $W_{1}$ contains one term and $W_{2}$ contains two terms.
\begin{equation}
A= i\frac{d}{dx} + W_{1}(x)= i\frac{d}{dx} + ix
\end{equation}
and 
\begin{equation}
B= i\frac{d}{dx} + W_{2}(x)=i\frac{d}{dx}+i x^{3} + i x^{5} 
\end{equation}
\begin{equation}
H^{(-)}=p^{2} -ixp-ix^{3}p-ix^{5}p-1-x^{4}-x^{6}
\end{equation}
and
\begin{equation}
H^{(+)}=p^{2} -ixp-ix^{3}p-ix^{5}p-3x^{2}-5x^{4}-x^{6}
\end{equation}
Few energy levels are plotted in figs-1,2 respectively.

\begin{figure}[htbp]
\centering 
\includegraphics[height=3.0in,width=5.0in]{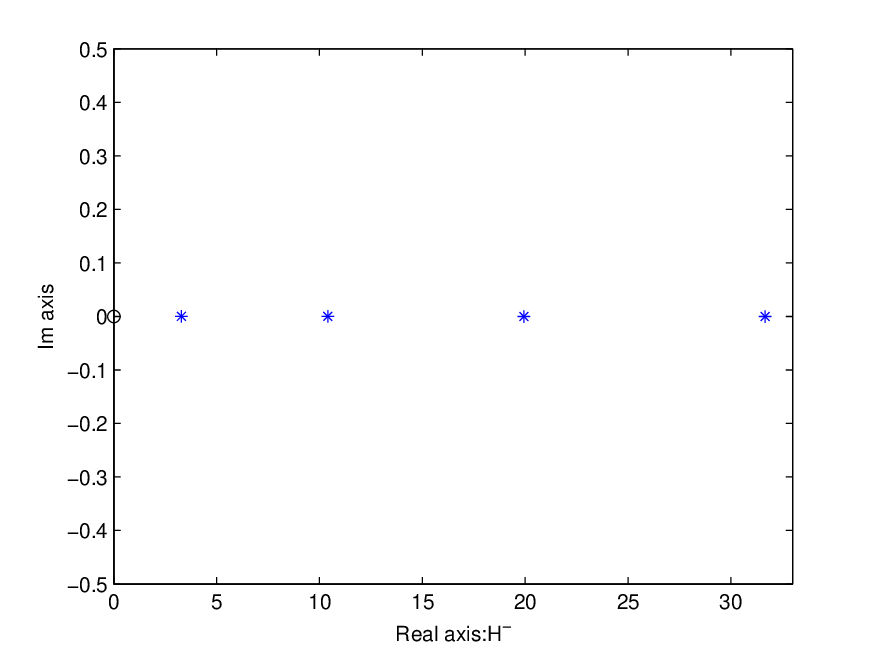}
\caption{ Spectra of $H^{(-)}$}
\includegraphics[height=3.0in,width=5.0in]{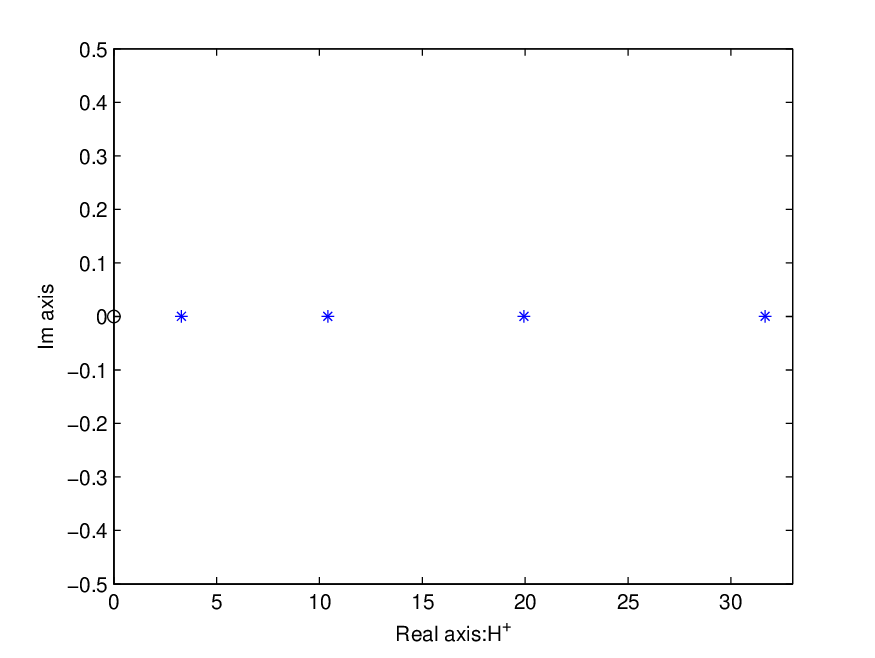}
\caption{ Spectra of $H^{(+)}$}
\end{figure}

\begin{table}[htbp]
Table-1 : Supersymmetric energy levels on complex potentials
\vspace{1.0cm}
\begin{tabular}{ c|c} \hline \hline
$H_{1}^{(-)}=p^{2}-ixp-ix^{3}p-ix^{5}p-1-x^{4}-x^{6}$&$ H_{1}^{(+)}=p^{2}-ixp-3x^{2}-ix^{3}p-ix^{5}p-6x^{4}-x^{6}$  \\ \hline
0            &  3.286 740 6              \\
3.286 740 6  & 10.403 315 5       \\ 
10.403 315 5 & 19.936 475 0    \\
19.936 475 0 & 31.666 346 2     \\ \hline
\end{tabular}
\end{table}

\begin{bf}
3.Method of calculation  
\end{bf}

Here we adopt matrix diagonalisation method[5] on solving the eigenvalue  relation  in complex space.

\begin{equation}
H |\Psi>= E |\Psi>
\end{equation}
where
\begin{equation}
|\Psi>=\sum A_{m}|\psi_{n}>
\end{equation}
with 
\begin{equation}
[p^{2}+x^{2}]\psi_{n}=(2n+1)\psi_{n}
\end{equation}

\begin{bf}
4.Conclusion
\end{bf}

In this paper, we develop supersymmetry for complex potentials using asymmetric formulation. Real spectra reported in table-1, confirms validity of the model proposed. Of course model potentials must be PT-symmetric in nature. Numerical results can easily extracted from the graphical plot.Other complex models can be generated[6] and will be reported elsewhere. Present model contains all the PT-terms  negative in nature and can hardly be seen in the literature.

\pagebreak
\begin{bf}

DATA AVAILABILITY : No additional data is required.

\vspace{0.1cm}

Conflict of interest : Author declares there is no conflict of interest of any kind.

\vspace{0.1cm}

Author's contribution: B.Rath: writing, reading, calculation,finalizing.

\end{bf}

\end{document}